\documentstyle[12pt,axodraw,epsf]{article}
\begin{document}
\begin{titlepage}
\setcounter{page}{0}  
\begin{center}
{\Large \bf CP Asymmetry for Inclusive Decay $B \to X_d +
\gamma$ in the Minimal Supersymmetric Standard Model}
\end{center}
\vspace{1.0cm}
\centerline{\large \bf H. H.~Asatryan, H. M.~Asatrian}
\vspace{0.5cm}
\centerline{\large Yerevan Physics Institute,
   2 Alikhanyan Br., 375036, Yerevan, Armenia}
\vspace{1.5cm}
\centerline{\bf ABSTRACT}
\vspace{0.5cm}
We study the inclusive rare decay $B\to X_d+\gamma$ in the
supergravity inspired Minimal Supersymmetric Standard Model
and compute the  CP-asymmetry in the decay rates. We show that
there exist two phenomenologically acceptable sets of SUSY parameters:
for one set the CP-asymmetry has the same (positive) sign as in the
Standard Model and lies in the range $(5-45)\%$ while for the other
set it is negative with values within $-(2-21)\%$.

\end{titlepage}

\section{Introduction}

The investigation of flavor-changing-neutral-currents (FCNC) induced 
B-meson decays will give an opportunity to search for new physics in 
the TeV region \cite{joan}.  In the Standard Model (SM) the FCNC
transitions are loop-induced. Therefore they are very sensitive 
to new particles mediating the respective transitions. 
The CP-asymmetry in the SM is caused by the single phase in the 
Cabibbo-Kobayashi-Maskawa (CKM) matrix and can be strongly affected 
by new CP-violating sources in  extensions of the Standard Model. 
The (mixing-induced) CP-symmetry violating effects were experimentally 
observed only in the decays of the neutral K-mesons \cite{pdg}. Hopefully the 
direct CP-asymmetry can be detected in the forthcoming explorations of 
the B-meson decays.

Supersymmetric models are good candidates to describe the
physics above the electroweak scale \cite{MSSM}. But even in the simplest
model with minimal particles content- the Minimal Supersymmetric Standard 
Model (MSSM)- there are many CP-violating phases and 
several new sources for FCNC processes which could lead to contradictions 
with available experimental data and consequently to strong constraints
on the parameters of the MSSM. 
Here we consider the model with supergravity
mediated soft breaking and grand unification. This scheme is assumed
to have an advantage of unifying  all particle interactions
including gravity and to give natural solution to the gauge
hierarchy problem \cite{MSSM}.

Penguin-induced FCNC decays have already been observed by the
CLEO collaboration \cite{CLEOrare1} and later by the ALEPH collaboration
\cite{ALEPHbsg}. In particular, the recent measurement of the branching
ratio of the inclusive decay $B\to X_s\gamma$ 
\cite{CLEOrare2}
\begin{equation}
\label{expbr}
BR(B\to X_s \gamma)=(3.15 \pm 0.35 \pm 0.32 \pm 0.26)\times10^{-4}
\end{equation}
is in good agreement with the SM prediction 
(see \cite{BURAS98, NEUBERT98}  and references therein) and
yields significant constraint on supersymmetric models.

The aim of this paper is to study the FCNC process
$B\to X_d \gamma$  for the supergravity inspired  MSSM. 
The inclusive decay $B\to X_d\gamma$  has not been  detected experimentally.
In the SM the ratio
\begin{displaymath}
R(d\gamma/s\gamma)\equiv
\frac{BR(B\to X_d\gamma)+BR(\overline{B}\to \overline{X_d}\gamma )}
{BR(B\to X_s\gamma)+BR(\overline{B}\to \overline{X_s}\gamma )}
\end{displaymath}
is known to lie in the range $0.017 \leq
R(d\gamma/s\gamma) \leq 0.074$ \cite{AAG98}. However the decay rate 
asymmetry of the  $B\to X_d\gamma$ decay is much larger than for 
the $B\to X_s\gamma$ decay and is possibly  detectable. The direct 
CP asymmetry (CP-asymmetry in the decay rates)
for the $B\to X_s\gamma$ decay in the SM and its nonsupersymmetric and
supersymmetric extensions was investigated in
\cite{SOARES91}-\cite{AYI97}. 
The CP-asymmetry for the $B\to X_d\gamma$ decay in the SM and 
left-right symmetric model  was 
calculated  in \cite{AAG98,AYI97}. The direct CP
asymmetry for the $B\to X_s\gamma$ decay
in the SM is found to be in the range -(0.4-1)\%, while for
$B\to X_d\gamma$ transition it varies in the interval (7-35)\%.

Our main goal is to calculate the direct CP-asymmetry for
$B\to X_d\gamma$ decay for  the supergravity inspired MSSM 
for the case when there are no extra 
CP violating SUSY parameters (i. e. there is only CKM-phase). 
We show that there exist two 
phenomenologically acceptable sets of SUSY parameters:  
for one set the CP-asymmetry, corresponding to the values  
$-0.1<\rho<0.4$, $0.2<\eta<0.5$ of the Wolfenstein parameters,
has the same (positive) sign as in the 
Standard Model and lies in the range $(5-45)\%$. For the other  set the 
CP asymmetry has negative sign and lies in the interval 
$-(2-21)\%$.

\section{The MSSM}

We begin with a short recapitulation of the model.
The general R-parity conserving superpotential is given by
\cite{MSSM,GDR}:
\begin{equation}
\label{superpot}
{\cal
W}=\sum_{i,j=gen}(-Y_{ij}^u\hat{u}_R^i\hat{H_u}\hat{Q^j}
+Y_{ij}^d\hat{d}_R^i\hat{H_d}\hat{Q^j}+
Y_{ij}^l\hat{l}^i_R\hat{H_u}\hat{L_j})+\mu\hat{H_u}\hat{H_d}.
\end{equation}
The corresponding supersymmetric Lagrangian  (including also the
kinetic terms and gauge-matter interactions), however leads
to the doubled particles spectrum which is unacceptable.
To avoid it,  one adds to the Lagrangian the so-called
soft breaking terms which explicitly violate the supersymmetry:
\begin{eqnarray}
\label{softbr}
\nonumber
&&-{\cal
L}_{gaugino}=\frac{1}{2}(M_1\tilde{b}\tilde{b}+M_2\sum_{a=1}^3
\tilde{w}^a\tilde{w}_a+M_3\sum_{a=1}^8 \tilde{g}^a\tilde{g}_a),\\
\nonumber
&&-{\cal
L}_{sferm}=\sum_{i=gen}(m^2_{\tilde{Q},i}\tilde{Q}_i^{+}\tilde{Q}_i+
m^2_{\tilde{L},i}\tilde{L}_i^{+}\tilde{L}_i+m^2_{\tilde{u},i}|
\tilde{u}_{R,i}|^2+
m^2_{\tilde{d},i}|\tilde{d}_{R,i}|^2+m^2_{\tilde{l},i}|\tilde{l}_{R,i}|^2)\\
&&-{\cal L}_{Higgs}=m^2_{H_u}H_u^{+}H_u+m^2_{H_d}H_d^{+}H_d+B\mu(H_u
H_d+h.c)\\
\nonumber
&&-{\cal L}_{tril.}=\sum_{i,j=gen}(A_{ij}^u Y_{ij}^u
\tilde{u}_{R,i}H_u \tilde{Q}_j+A_{ij}^d Y_{ij}^d
\tilde{d}_{R,i}H_d \tilde{Q}_j+A_{ij}^l Y_{ij}^l
\tilde{l}_{R,i}H_u \tilde{L}_j+h.c.).
\end{eqnarray}
The expressions (\ref{superpot}), (\ref{softbr}) determine the MSSM
in the most general (unconstrained) form. In addition to the 19 parameters
of the SM the MSSM contains another 105 
parameters. The total number of the parameters can be reduced 
assuming that the MSSM
parameters obey specific (universal) boundary conditions at the
GUT scale. These assumptions are natural in the scenario where
the SUSY-breaking occurs via gravitational interaction with the
hidden sector. Without discussing the details of the hidden
sector, the actual scale and scheme of the Grand Unification
we  impose the following conditions:
\begin{eqnarray}
\label{condit}
\nonumber
&&\alpha_1(M_{GUT})=\alpha_2(M_{GUT})=\alpha_3(M_{GUT})\equiv\alpha_U\\
\nonumber
&&M_1(M_{GUT})=M_2(M_{GUT})=M_3(M_{GUT})\equiv m_{1/2}\\
\nonumber
&&M_{\tilde{Q}}(M_{GUT})=M_{\tilde{u}_R}(M_{GUT})=M_{\tilde{d}_R}(M_{GUT})=
M_{\tilde{L}}(M_{GUT})=M_{\tilde{l}_R}(M_{GUT})\equiv
m_0\\
&&M_{H_u}(M_{GUT})=M_{H_d}(M_{GUT})\equiv m_H\\
\nonumber
&&A_u(M_{GUT})=A_d(M_{GUT})=A_l(M_{GUT})\equiv A_0.
\end{eqnarray}

To obtain the values of the running masses and couplings relevant
for our discussion, the renormalization group equations (RGE) have to be
solved performing the evolution from the scale $M_{GUT}$  to the 
electroweak scale.

In general, one must ensure that electroweak symmetry breaking
(EWSB) occurs. For this the following two necessary conditions
must be satisfied:\\
\begin{equation}
\label{EWSB}
\frac{1}{2}M_{Z}^2=\frac{\overline{m}_1^2-\overline{m}_2^2
tan^2\beta}{tan^2\beta-1}
\quad \quad \quad
sin2\beta=\frac{-2\overline{m}^2_3}{\overline{m}_1^2+\overline{m}_2^2},
\end{equation}
where
\begin{eqnarray}
&&\overline{m}_1^2=m_{H_d}^2+\mu^2+rad.corr.,\nonumber\\
&&\overline{m}_2^2=m_{H_u}^2+\mu^2+rad.corr.,\\
&&\overline{m}_3^2=-B\mu\nonumber
\end{eqnarray}
and $\beta$ is defined as usual. We consider the set of quantities
\begin{equation}
\label{param}
m_0,m_{1/2},A_0,\beta,sign(\mu)
\end{equation}
as independent parameters, while B and $\mu$ are determined through 
Eq. (\ref{EWSB}).

\section{The effective Hamiltonian}

We study the $B\to X_{s(d)}+\gamma$ decay using  the effective theory
obtained by integrating out the heavy degrees of freedom, which are the
W-boson, t-quark and all the SUSY-particles.
The effective Hamiltonian for the $B\to X_{s(d)}\gamma$   decay 
%in the SM (and also in the MSSM in leading order)  
has the following form 
(keeping operators up to the dimension 6):
\begin{eqnarray}
\label{heffbsgcp}
&&{\cal H}_{eff}(b \to s(d) \gamma (+g))  =
-  \frac{4G_F}{\sqrt{2}}
\left\{ \lambda_{t}^{s(d)}  \sum_{i=1}^8 C_i(\mu) O_i(\mu) \right. \\
\nonumber
&& \left.
-\lambda_u^{s(d)}C_{2}(\mu)(O_{2u}(\mu)-O_{2}(\mu))- 
\lambda_u^{s(d)}C_{1}(\mu)(O_{1u}(\mu)-O_{1}(\mu))\right \}
\end{eqnarray}
where the operator basis $O_i$ and the Wilson coefficients
$C_i$ can be seen elsewhere \cite{BURAS98,Misiak96}. 
The quantities $\lambda_t^{s(d)}\equiv
V_{tb}V_{ts(d)}^*$, $\lambda_u^{s(d)}\equiv V_{ub}V_{us(d)}^{*}$
are the relevant CKM factors. In the Wolfenstein
parametrization these factors can be expressed in the following form
\cite{AAG98}:
\begin{eqnarray}
&&\lambda_u^d=A\lambda^3(\bar{\rho}-i\bar{\eta})
\quad \quad
\lambda_t^d=A\lambda^3(1-\bar{\rho}+i\bar{\eta})\\
\nonumber
&&\lambda_u^s=A\lambda^4(\rho-i\eta)
\quad \quad
\lambda_t^s=-A\lambda^2(1-\frac{\lambda^2}{2}+\lambda^2(\rho-i\eta))
\end{eqnarray}
with $\overline{\rho}=\rho (1-\lambda^2/2),\quad
\overline{\eta}=\eta ( 1-\lambda^2/2)$.

For the supersymmetric case an additional 
set of operators can exist. However, for our case with universal 
boundary conditions (\ref{condit}) we can consider that the set of 
operators in SUSY is the same as in SM \cite{unpub}. 
The  processes $\overline{B}\to \overline{X_{s(d)}}\gamma$ can be well described by the
partonic level transitions $b\to s(d)\gamma$ and the leading 
corrections can be systematically obtained using the heavy quark 
effective theory (see \cite{BURAS98, NEUBERT98, AAG98} and references
therein).
The calculation of the corresponding partonic transitions in a
given order on $\alpha_s(m_b)$ includes the following three steps
\cite{BURAS}:
\begin{itemize}
\item[(i)] The Wilson coefficients $C_i$ at the scale $M_W$ must
be calculated  "matching" the effective and full theories;
\item[(ii)] Then RGE must be used to obtain the Wilson
coefficients at the scale $\mu\sim m_b$. This step requires
the knowledge of the anomalous dimension matrix;
\item[(iii)] The matrix elements of the operators $O_i$ for the
processes $b\to s(d)\gamma$ and $b\to s(d)\gamma g$ have to
be calculated.
\end {itemize}
The crucial point here is that for considering case 
SUSY effects enter only through step (i). 
For the SM two-loop matching has been performed in
\cite{Yao94,Greub97}. The full two-loop matching for the SM
extension with two Higgs doublets (2HDM) can be found in
\cite{new,BG98}. To the best of our knowledge the complete matching
for  MSSM has not been done yet (see, for example \cite{ciuch}).
The anomalous dimension matrix in the next-to-leading order
is known from \cite{Misiak96}. We give here the leading order relations
between the Wilson coefficients at the high and low scales:
\begin{eqnarray}
\label{wilsoncoeff}
\nonumber
&&C_2(\mu)=\frac{1}{2}(\eta^{-\frac{12}{23}}+\eta^{\frac{6}{23}})\\
&&C_{7}^{eff}(\mu)=\eta^{\frac{16}{23}}C_7(M_W)+\frac{8}{3}
(\eta^{\frac{14}{23}}-\eta^{\frac{16}{23}})
C_8(M_W)+\sum_{i=1}^8 h_i\eta^{a_i}\\
&&C_{8}^{eff}(\mu)=\eta^{\frac{14}{23}}C_8(M_W)+\sum_{i=1}^8
\overline{h_i}\eta^{a_i},
\nonumber
\end{eqnarray}
where $\eta=\alpha_s(M_W)/\alpha_s(\mu)$ and the
full next-to-leading order result is given in  \cite{Misiak96}. 
Although the complete expressions for the $C_7(M_W)$ and $C_8(M_W)$ in the
MSSM are still   unknown, we use the next-to-leading order approximation
for the anomalous dimension matrix, operator matrix elements  and
the matching conditions for the SM contribution.
The virtual $O(\alpha_s)$ corrections to the matrix
elements for the decay $b\to s\gamma$ are calculated in
\cite{GHW96}.
As to the Bremsstrahlung corrections, which are presented in
\cite{ag95,Pott95}, we choose the scheme from
\cite{Misiak96} with a low-energetic cutoff on the photon energy.

The relevant operators are
\begin{eqnarray}
\label{operators}
O_{2}(\mu)&=&(\bar{c}_{L\alpha}\gamma_{\mu} b_{L\alpha})
(\bar{s}_{L\beta}\gamma^{\mu} c_{L\beta}) \hspace{1.4cm}
O_{2u}(\mu)=(\bar{u}_{L\alpha}\gamma_{\mu} b_{L\alpha})
(\bar{s}_{L\beta}\gamma^{\mu} u_{L\beta})\\
\nonumber
O_{7}(\mu)&=&\frac{e}{16\pi^2}m_{b}(\mu)(\bar{s}_{L}\sigma_{\mu
\nu}b_{R}) F^{\mu \nu} \quad  \quad
O_8(\mu)=\frac{g}{16\pi^2}m_b(\mu)(\bar{s}_{L}T^{a}
\sigma_{\mu \nu}b_{R})G^{a \mu \nu},
\end{eqnarray}
where $T^a$ are the generators of color SU(3) group.
The contribution of other operators in (\ref{heffbsgcp})
is suppressed due to the smallness of their Wilson coefficients.

It is convenient to write the decay rate for the process
$\overline{B}\to \overline{X_{s(d)}}\gamma$
introducing the quantities $D_{t}$, $D_{u}$, $D_{r}$, $D_{i}$
\cite{AAG98} as follows:
\begin{eqnarray}
\label{brd}
BR(\overline{B} \to
\overline{X_{s(d)}}\gamma)&=&\frac{|\lambda_t^{s(d)}|^2}
{|V_{cb}|^2}D_{t}+
\frac{|\lambda_u^{s(d)}|^2}{|V_{cb}|^2}D_{u}+\\
\nonumber
&&+\frac{Re(\lambda^{*s(d)}_t\lambda_u^{s(d)})}{|V_{cb}|^2}D_{r}+
\frac{Im(\lambda^{*s(d)}_t\lambda_u^{s(d)})}{|V_{cb}|^2}D_{i}
\end{eqnarray}
The corresponding expression for the $B \to X_{s(d)}\gamma$ decay rate can
be obtained from (\ref{brd}) by changing the sign of the term proportional
to  $Im(\lambda^{*s(d)}_t\lambda_u^{s(d)})$.

We define the direct CP asymmetry for the $B\to X_{s(d)} \gamma$ decay as:
\begin{equation}
\label{acp1}
a_{CP}(B\to X_{s(d)}\gamma)=\frac{\Gamma(B\to X_{s(d)} \gamma)-
\Gamma(\overline{B}\to \overline{X_{s(d)}} \gamma)}
{\Gamma(B\to X_{s(d)} \gamma)+
\Gamma(\overline{B}\to \overline{X_{s(d)}} \gamma)}.
\end{equation}
The CP asymmetry is then:
\begin{equation}
\label{acp0}
a_{CP}(B\to X_{s(d)}\gamma)=-\frac{Im(\lambda_u^{s(d)}
\lambda_t^{*s(d)})D_{i}}
{|\lambda_t^{s(d)}|^2D_{t}^{(0)}}
\end{equation}
The quantities $D_a$, a=t,u,i,r are  the same for both decays
\footnote{Note that there is a weak dependence on the fraction
$m_b/m_s$ when we use low energetic cutoff $\delta$ for the
bremsstrahlung corrections, restricting the photon energy spectrum
by the condition $E_{\gamma}<m_B(1-\delta)/2$. For numerical estimates
we take $\delta=0.99$.}
(for the SM they are given in \cite{AAG98}).
Note that $D_{t}^{(0)}$ is calculated in  leading logarithmic 
approximation
using one loop expression for $\alpha_s$ \cite{AAG98}.

As usual we express the
branching ratio  $BR(B\to X_{s(d)}+\gamma)$ in terms of the measured
semileptonic  branching ratio $BR (B \to X\ell \nu_\ell)$:
\begin{equation}
\label{brbsgsm}
BR(B\to X_{s(d)}\gamma) = \frac{\Gamma(B\to X_{s(d)}\gamma)}{\Gamma_{sl}}
\, BR (B \to X\ell \nu_\ell).
\end{equation}
The expression for the $\Gamma_{sl}$ (including QCD corrections)
can be found  in \cite{Cabibbo}.

\section{$B\to X_d+\gamma$ decay rate and CP-asymmetry in the
 MSSM}

In the MSSM in addition to the $W^{\pm}$
mediated diagrams Fig. 1a, there are several new sources
for the $b\to s$ transition (Fig. 1b, Fig. 2).
It is known that chargino and charged Higgs contribution are 
dominant while neutralino and
gluino contributions are small and can be neglected \cite{Okada98}.
The Wilson coefficients $C_7(M_W)$ and $C_8(M_W)$  in the SM was
calculated in \cite{Yao94}. For our case we have:
\begin{equation}
C_{7(8)}=C^{SM}_{7(8)}+C^{H^{\pm}}_{7(8)}+C^{\chi^{\pm}}_{7(8)}.
\end{equation}
The contributions from $H^{\pm}$, $\chi^{\pm}$
exchange diagrams are given in \cite{Oshimo98,Oshi}.
The Wilson coefficient $C_2(M_W)$ in the leading order
is the same as in the SM.

In general the CP-asymmetry is determined by the complex couplings in
the Lagrangian. In the MSSM these can be $A_0$ and $\mu$. 
However the existence of new physical phases can lead to large values for 
the electric dipole
moment (EDM) for the neutron and leptons. Thus experimental restrictions on
EDM limit possible values of corresponding phases. In
our case these phases enter through mass matrices of top-squarks
and charginos \cite{GDR}:
\begin{eqnarray}
\label{masmat}
\nonumber
&&M_{\tilde{t}}^2=\left( \begin{array}{cc}
\tilde{M}_{uL}^2+m_t^2&m_t(A^{*}-cot\beta\mu)\\
m_t(A-cot\beta\mu^{*})&\tilde{M}_{uR}^2+m_t^2
\end{array} \right)\\
&&   \\
\nonumber
&&M_{\chi^{\pm}}=\left( \begin{array}{cc}
\tilde{M_2}&\sqrt{2}cos\beta M_W \\
\sqrt{2}sin\beta M_W&\mu
\end{array} \right).
\end{eqnarray}
The EDM constraints allow the phase $arg(\mu)$ to have non-negligible
values only if the masses of squarks are of order 1 TeV 
\cite{IN98}-\cite{Nihei}.
As shown in \cite{Okada98} for our case with universal
scalar masses and trilinear couplings, the phase of the
$A_0$ is small: even if it has large value at the GUT scale, due to the 
RGE evolution of the $A_0$ from the $M_{GUT}$ down
to the electroweak scale  it becomes small. This is not valid for the 
case when squarks masses are of order of 1TeV, but in that case the
contribution of diagrams in Fig. 2 is negligible.
The resulting phase cannot lead  to
significant deviations from the SM predictions. In contrast to 
\cite{CHH98} and
\cite{Oshimo98} (see also \cite{NK98}) where phenomenological approach is 
carried on,
the CP-asymmetry for  $B \to X_s\gamma$ decay in our scheme
is no more enhanced by SUSY effects and so
remains small and unobservable.

Our aim is to obtain limits for the $B \to X_d \gamma$ decay rate
asymmetry in the SUGRA inspired MSSM. We will use the experimentally
obtained range for $B \to X_s \gamma$ decay rate (\ref{expbr}) which puts 
strong restrictions on SUSY parameters. As mentioned, we will consider
the case when there are no CP violating SUSY parameters, i.e. the
mass matrices in (\ref{masmat}) are real.

In \cite{AAG98} one of the main targets of study is the ratio
$R(d\gamma /s\gamma )$. As it is obvious from Eq.(\ref{brd}), where the 
term proportional  to $D_t$ is dominant, in the MSSM
this ratio is approximately the same as in the SM. Thus the only significant 
deviation from the SM which can be found studying
the $B\to X_d \gamma$ (and $\overline{B}\to \overline{X_{d}} \gamma$) 
decay rate is the
CP-asymmetry. A sufficiently accurate measurement
of the CP asymmetry will require large statistics and will not be achieved 
in the near future.  It will 
be easier to determine the sign of the CP-asymmetry, which as is shown 
below, can be different from that of the SM.\\
%
%
%******************************************************************
%
For the CP-asymmetry in the
$B\to X_d\gamma$ decay we have \cite{NK98}:
\begin{equation}
\label{acp}
a_{CP}(B\to X_d \gamma)=\frac{8z}{27}\alpha_s\bar{\eta} 
\frac{3(v(z)+b(z,\delta))C_2C_7-
b(z,\delta)C_2C_8}
{C_7^2\left [ (1-\bar{\rho})^2+\bar{\eta}^2 \right ]}
\end{equation}
where $z=m_c^2/m_b^2$, the expressions for $v(z)$ and $b(z,\delta)$
can be found in \cite{NK98}.
The first term in the numerator (\ref{acp})
is dominant thus the sign of the $C_7$ determines the sign of 
the $a_{CP}(B\to X_d \gamma)$.

We scan the parameter space of the MSSM assuming that
$m_0,m_{1/2}<1000GeV$, $-4000GeV<A_0<4000GeV$, $1.5<tan\beta<50$.
Starting with a given set of values of the parameters
(\ref{param}) we use the RGE to
perform the evolution from $M_{GUT}$ down to  low energy scale. Also the
existence of appropriate EWSB must be guaranteed. For these
two steps we use the program SUSPECT from GDR group
\cite{GDR} which gives numerical solution of RGE.
We take into account the experimental restrictions for sparticle
and Higgs masses \cite{chOPAL}-\cite{higgs}. 
Of particular importance are the restrictions 
on the stop quark and chargino masses:
\begin{equation}
\label{susymass}
m_{\chi^{\pm}}\geq 91GeV \quad \quad m_{\tilde{t}}\geq 82GeV.
\end{equation}

The experimental result (\ref{expbr}) for $B\to X_s \gamma$ decay rate
puts strong constraints on the absolute value of the
Wilson coefficient $C_7(m_b)$. However, as already mentioned in
\cite{Okada97} the sign of $C_7(m_b)$ remains undetermined. 
The sign of $C_7(m_b)$ in the MSSM can be opposite to that in 
the SM if the contributions coming from the diagrams Fig. 1b, Fig. 2 
will have the required sign and absolute value. In the so-called
type II two Higgs doublet model
and MSSM charged Higgs contribution has the same (negative) sign
as $C_7^{SM}(m_b)$ \cite{BG98}. In the MSSM there exist
values of the SUSY parameters for which  chargino contribution is
positive and large enough to have $C_7(m_b)\sim -C_7^{SM}(m_b)$.

As a result of numerical calculations we find two sets of
SUSY parameters (\ref{param}) satisfying  to the mentioned above conditions for
sparticle masses and decay rate. For one set  $C_7(m_b)$ is negative
(as in the Standard Model) while for the other set its sign is positive.
We do not present here complete restrictions for these parameters, 
corresponding to these sets as they are rather complicated.
Instead we present some qualitative results.
We find that the $C_7(m_b)$ can has positive sign for large values
of $\tan\beta$ ($\tan \beta
\geq 25$), $sign(\mu)<0$ and relatively low values of the $m_{1/2}$
($m_{1/2}\leq 160GeV$)
in the large range of  the parameter space. Besides, for the positive sign
of  the $ C_7(m_b)$ the masses of the light stop and chargino are
relatively low  ($m_{\chi^{\pm}}\leq 120GeV$, $m_{\tilde{t}}\leq 180GeV$).

Now we discuss the functions $D_a$, a=t,i which determine the
$B\to X_{s(d)} \gamma$ decay asymmetry.
In Table 1  we present the functions $D_t^{(0)}$ and
$D_i$ evaluated for  the central values of the parameters 
$\alpha_s=0.118$, $m_c/m_b=0.29$, $m_t=175GeV$, $m_b=4.8GeV$. 
The Wilson coefficient $C_7(m_b)$  has the same (negative) sign as in 
the SM. For the scale parameter and branching ratio we take:  
$\mu=2.5,5,10GeV$, $BR(B\to X_s\gamma)\times 10^4=2.61,3.15,3.69$.  
In Table 2  we give the
values of the same functions for $C_7(m_b)>0$. For both cases for the
given value of the branching ratio of the $B\to X_s\gamma$ decay
$D_t^{(0)}$ and $D_i$ do not vary more than $\sim$ 1\% with the variation
of the SUSY parameters. The reason is that the branching ratio of
$B\to X_s\gamma$ decay together with the sign of the $C_7(m_b)$
determines the Wilson coefficient $C_7(m_b)$
with the precision of $\sim$ 1\%, as $C_2$ doesn't vary with the change of the
SUSY parameters and $C_8(m_b)$ is small. Thus, we will not give the 
dependence of the $D_a$ on the SUSY parameters. Note that the inclusion 
of the next-to-leading order corrections,
corresponding to the SUSY particles contribution can slightly change 
the situation, leading to small change in the dependence of $D_t$, $D_i$ 
on the SUSY parameters. 

For the numerical estimates of the CP-asymmetry and
the decay rate we take $-0.1<\rho<0.4$, $0.2<\eta<0.5$. For the decay
$B\to X_s \gamma$ the direct CP-asymmetry is too small to be
measurable: for $2.61<BR(B\to X_s\gamma)<3.69$ it varies  in
the range -(0.3-1.3)\% for $C_7(m_b)<0$ and (0.1-0.7)\% for $C_7(m_b)>0$.

We now consider the decay $B\to X_d\gamma$. As already has been
mentioned the deviation of the ratio $R(d\gamma/s\gamma)$ from its SM 
value is small. This ratio is given by \cite{AAG98}
\begin{displaymath}
\nonumber
R(d\gamma/s\gamma)=\lambda^2\left[ 1+\lambda^2(1-2\overline{\rho})\right]
\left[ (1-\overline{\rho})^2+\overline{\eta}^2+
\frac{D_u}{D_t}(\overline{\rho}^2+
\overline{\eta}^2)+\frac{D_r}{D_t}(\overline{\rho}(1-\overline{\rho})
-\overline{\eta}^2) \right]
\end{displaymath}
and in the first approximation doesn't depend on the values of  $D_a$.
The numerical calculations show, that for any $\rho$ and $\eta$ the
deviation of the $R(d\gamma/s\gamma)$ due to the variation of  SUSY 
parameters doesn't exceed $\sim$ 1\%.

Let us proceed to the CP-asymmetry in the $B\to X_d\gamma$ decay.
Using the values of the functions  $D_t^{(0)}$ and $D_i$ from
Table 1 and 2 we can obtain the range for the direct CP-asymmetry 
$a_{CP}(B\to X_d \gamma)$ for two sets of the SUSY parameters.  
When the sign of
the $C_7(m_b)$ coincides with that in the SM the asymmetry is positive
and lies in
the range $5\%<a_{CP}(B\to X_d \gamma)<45\%$. When the sign  of 
$C_7(m_b)$ is positive
then the CP-asymmetry is negative and lies in the range
$-21\%<a_{CP}(B\to X_d \gamma)<-2\%$. The absolute value of the CP 
asymmetry reaches its maximum  and minimum values for $\rho=0.4$, $\eta=0.5$,
$Br(B\to X_s\gamma)=2.61\times 10^{-4}$ and $\rho=-0.1$,
$\eta=0.2$, $Br(B\to X_s\gamma)=3.69\times 10^{-4}$ respectively.
Thus for both cases ($C_7(m_b)<0, C_7(m_b)>0$) the CP-asymmetry for
the $B\to X_d \gamma$ decay is large enough and is expected to be
measurable at high luminosity B factories. The observation of
the CP-asymmetry with negative sign will be a clear manifestation
of the physics beyond the SM and in particular of the
Supersymmetry (see also \cite{AYI97}).

We investigate also the dependence of the CP-asymmetry on the
Wolfenstein parameters $\rho$, $\eta$ and the $B\to X_s \gamma$
decay rate. For the given values of the branching
ratio of the $B\to X_s\gamma$ decay and the SM parameters $\rho$ and $\eta$
the decay rate asymmetry $a_{CP}(B\to X_d \gamma)$ does not vary more 
than $\sim$ 1\% with the variation of  SUSY parameters.
In Fig. 3 we present the direct CP-asymmetry for
the decay $B\to X_d\gamma$  as a function of the branching ratio
of the $B\to X_s\gamma$ decay for $\rho=0.11$ and $\eta=0.20,0.33,0.50$. 
The absolute value CP asymmetry for both cases ($C_7(m_b)<0, C_7(m_b)>0$)
is largest for  the largest value of $\eta$. It grows with decrease of
the branching ratio of $B\to X_s\gamma$ decay. The dependence of the 
CP-asymmetry on the $B\to X_s\gamma$ decay branching ratio is stronger 
for the case $C_7(m_b)<0$.  In Fig. 4 we present the direct CP-asymmetry for
the decay $B\to X_d\gamma$  as a function of the branching ratio
of $B\to X_s\gamma$ decay for $\eta=0.33$ and $\rho=-0.1,0.11,0.4$. 
The absolute value CP asymmetry for both cases ($C_7(m_b)<0, C_7(m_b)>0$)
is largest for  the smallest value of $\rho$.   Note that in Fig. 3,4 we 
use the value $\mu=2.5$GeV for the scale parameter. For this value NLL 
corrections to the decay rate become minimal \cite{AAG98}. In
Table 3 we illustrate the
dependence of the CP-asymmetry on the scale parameter $\mu$: we give 
the values of the $a_{CP}(B\to X_d\gamma)$
for $\rho=0.11$, $\eta=0.33$ and for $BR(B\to X_s \gamma)\cdot 10^4 
=2.61,3.15,3.69$,
$\mu=2.5,5,10GeV$. Note that the scale dependence of the CP asymmetry is
stronger for the case $C_7>0$.

In conclusion, we have presented the theoretical estimates for the  direct
CP-asymmetry $a_{CP}(B \to X_d \gamma )$ in the MSSM when the all
SUSY parameters are real. We found two phenomenologically acceptable
sets of SUSY parameters. For one CP-asymmetry has  the same (positive) 
sign as in the Standard Model and for the other set its
sign is opposite. For both cases CP asymmetry is large enough to be 
eventually observed. 
A more detailed analysis of the CP-asymmetry for $B \to X_d \gamma$ decay 
including the case when there exist new sources of CP-violation in
the MSSM will be given elsewhere.

%\newpage
\vspace{1cm}
{\large \bf Acknowledgments}
\vspace{0.4cm}       

We thank D. Wyler and Ch. Greub for discussions. This work was partially
supported by INTAS under  Contract INTAS-96-155.

\vspace{3cm}
\begin{center}
{\bf \large Figure Captions}
\end{center}
\vspace{0.5cm}

Figure 1: Feynman graphs mediated by the exchange of (a) W-bosons
and (b) charged Higgs bosons $H^{\pm}$. Possible places of
photon emission are labeled with a  cross.

Figure 2: Feynman graphs mediated by the exchange of SUSY
particles: (a) charginos $\chi^{\pm}$,
(b) gluinos $\tilde{g}$ and (c) neutralinos $\chi^0$. Possible places of
photon emission are labeled with a cross.

Figure 3: The dependence of the $a_{CP}(B\to X_d\gamma)$ from 
$BR(B\to X_s\gamma)$ for the values $\eta=0.20,0.33,0.50$ and
$\rho=0.11$, $\mu=2.5GeV$.
The CLEO bounds are shown.

Figure 4: The dependence of the $a_{CP}(B\to X_d\gamma)$ from 
$BR(B\to X_s\gamma)$ for the $\rho=-0.10,0.11,0.40$ and
$\eta=0.33$, $\mu=2.5GeV$. The CLEO bounds are shown.

\newpage

\begin{center}
Table 1.\\

Values of the functions $D_t^{(0)}$ and $D_i$
for  $\mu=2.5,5,10GeV$; $BR(B\to X_s\gamma)\times 10^4=2.61,3.15,3.69$
and $C_7<0.$
\end{center}
\vspace{0.5cm}
\begin{tabular}{|l|c|c|c|c|} \hline
&$\mu =2.5GeV$&$\mu =5GeV$&$\mu =10GeV$&$Br(B\to X_s\gamma)\times 10^4$\\ 
\hline
$D_t^{(0)}/\lambda^4$&0.095&0.076&0.064&2.61\\ \hline
$D_t^{(0)}/\lambda^4$&0.120&0.099&0.086&3.15\\ \hline
$D_t^{(0)}/\lambda^4$&0.147&0.124&0.109&3.69\\ \hline
$D_i/\lambda^4$&0.052&0.036&0.026&2.61\\ \hline
$D_i/\lambda^4$&0.057&0.040&0.030&3.15\\ \hline
$D_i/\lambda^4$&0.063&0.044&0.033&3.69\\ \hline
\end{tabular}
\vspace{1.5cm}
\begin{center}
Table 2.\\

Values of the functions $D_t^{(0)}$ and $D_i$
for  $\mu=2.5,5,10GeV$; $BR(B\to X_s\gamma)\times 10^4=2.61,3.15,3.69$
and $C_7>0$.
\end{center}
\vspace{0.5cm}
\begin{tabular}{|l|c|c|c|c|} \hline
&$\mu =2.5GeV$&$\mu =5GeV$&$\mu =10GeV$&$Br(B\to X_s\gamma)\times 10^4$\\ 
\hline
$D_t^{(0)}/\lambda^4$&0.193&0.364&0.611&2.61\\ \hline
$D_t^{(0)}/\lambda^4$&0.228&0.412&0.677&3.15\\ \hline
$D_t^{(0)}/\lambda^4$&0.261&0.459&0.742&3.69\\ \hline
$D_i/\lambda^4$&-0.053&-0.061&-0.066&2.61\\ \hline
$D_i/\lambda^4$&-0.059&-0.065&-0.070&3.15\\ \hline
$D_i/\lambda^4$&-0.063&-0.069&-0.074&3.69\\ \hline
\end{tabular}

\vspace{1.5cm}
\begin{center}
Table 3.\\

Values of the $a_{CP}(B\to X_d \gamma)$  
for $\rho=0.11$, $\eta=0.33$ and for $BR(B\to X_d \gamma) \times 10^4 
=2.61, 3.15,3.69$; $\mu=2.5,5,10GeV$.
\end{center}
\vspace{0.5cm}
\begin{tabular}{|l|c|c|c|c|} \hline
&$\mu =2.5GeV$&$\mu =5GeV$&$\mu =10GeV$&
$Br(B\to X_s\gamma)\times 10^4$\\ \hline
$a_{CP}(B\to X_d \gamma)$, $C_7<0$&0.20&0.17&0.15&2.61\\ \hline
$a_{CP}(B\to X_d \gamma)$, $C_7<0$&0.17&0.15&0.13&3.15\\ \hline
$a_{CP}(B\to X_d \gamma)$, $C_7<0$&0.16&0.13&0.11&3.69\\ \hline
$a_{CP}(B\to X_d \gamma)$, $C_7>0$&-0.10&-0.061&-0.040&2.61\\ \hline
$a_{CP}(B\to X_d \gamma)$, $C_7>0$&-0.095&-0.058&-0.038&3.15\\ \hline
$a_{CP}(B\to X_d \gamma)$, $C_7>0$&-0.088&-0.055&-0.037&3.69\\ \hline
\end{tabular}

\newpage
\begin{figure}

\begin{center}
\begin{picture}(350,100)(0,0)
\SetWidth{0.8}

\ArrowLine(0,70)(40,70)\ArrowLine(40,70)(80,70)
\ArrowLine(80,70)(120,70)\ArrowLine(120,70)(160,70)
\PhotonArc(80,70)(40,90,180){2}{4}
\PhotonArc(80,70)(40,0,90){2}{4}
\Text(80,70)[]{\boldmath $\times$}
\Text(0,65)[lt]{$b_L$}\Text(160,65)[rt]{$s_L(d_L)$}
\Text(80,65)[t]{$u,c,t$}\Text(80,110)[]{\boldmath $\times$}
\Text(47,100)[r]{$W^{-}$}
\Text(30,70)[]{\boldmath $\times$}
\Text(130,70)[]{\boldmath $\times$}

\Text(80,25)[]{\bf (a)}

\ArrowLine(190,70)(230,70)\ArrowLine(230,70)(270,70)
\ArrowLine(270,70)(310,70)\ArrowLine(310,70)(350,70)
\DashCArc(270,70)(40,90,180){4}\DashCArc(270,70)(40,0,90){4}
\Text(200,65)[lt]{$b_L$}\Text(350,65)[rt]{$s_L(d_L)$}
\Text(270,70)[]{\boldmath $\times$}
\Text(270,110)[]{\boldmath $\times$}
\Text(220,70)[]{\boldmath $\times$}
\Text(320,70)[]{\boldmath $\times$}
\Text(237,100)[r]{$H^{+}$}
\Text(270,25)[]{\bf (b)}
\Text(270,65)[t]{$u,c,t$}

\end{picture}
\end{center}
\vspace{-1.cm}
\caption{Feynman graphs mediated by the exchange of (a)
W-bosons and (b) charged Higgs bosons $H^{\pm}$. Possible places of
photon emission are labeled with a cross.}\label{f1}
\end{figure}
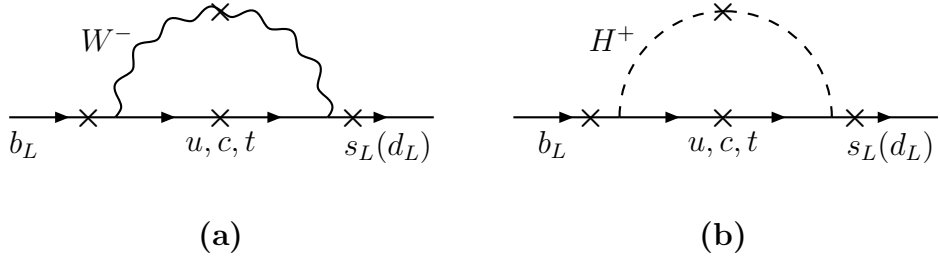

\begin{figure}
\begin{center}
\begin{picture}(350,150)(0,0)
\SetWidth{0.8}

\ArrowLine(0,130)(40,130)\DashLine(40,130)(80,130){3}
\DashLine(80,130)(120,130){3}\ArrowLine(120,130)(160,130)
\ArrowArcn(80,130)(40,180,90)
\ArrowArcn(80,130)(40,90,0)
\Text(80,130)[]{\boldmath $\times$}
\Text(0,125)[lt]{$b_L$}\Text(160,125)[rt]{$s_L(d_L)$}
\Text(80,125)[t]{$\tilde{u},\tilde{c},
\tilde{t}$}\Text(80,170)[]{\boldmath $\times$}
\Text(47,160)[r]{$\chi_{1,2}^{\pm}$}
\Text(30,130)[]{\boldmath $\times$}
\Text(130,130)[]{\boldmath $\times$}

\Text(80,95)[]{\bf (a)}

\ArrowLine(190,130)(230,130)\DashLine(230,130)(270,130){3}
\DashLine(270,130)(310,130){3}\ArrowLine(310,130)(350,130)
\ArrowArcn(270,130)(40,180,90)\ArrowArcn(270,130)(40,90,0)
\Text(200,125)[lt]{$b_L$}\Text(350,125)[rt]{$s_L(d_L)$}
\Text(270,130)[]{\boldmath $\times$}
\Text(220,130)[]{\boldmath $\times$}
\Text(320,130)[]{\boldmath $\times$}
\Text(237,160)[r]{$\tilde{g}$}
\Text(270,95)[]{\bf (b)}
\Text(270,125)[t]{$\tilde{d},\tilde{s},\tilde{b}$}

\ArrowLine(80,30)(120,30)\DashLine(120,30)(160,30){3}
\DashLine(160,30)(200,30){3}\ArrowLine(200,30)(240,30)
\ArrowArcn(160,30)(30,180,90)\ArrowArcn(160,30)(30,90,0)
\Text(90,25)[lt]{$b_L$}\Text(250,25)[rt]{$s_L(d_L)$}
\Text(160,30)[]{\boldmath $\times$}
\Text(110,30)[]{\boldmath $\times$}
\Text(210,30)[]{\boldmath $\times$}
\Text(127,60)[r]{$\tilde{\chi^0}$}
\Text(160,25)[t]{$\tilde{d},\tilde{s},\tilde{b}$}
\Text(160,-5)[]{\bf (c)}
\end{picture}
\end{center}

\vspace{0.cm}
\caption{ Feynman graphs mediated by the exchange of SUSY
particles: (a) charginos $\chi^{\pm}$,
(b) gluinos $\tilde{g}$ and (c) neutralinos $\chi^0$ Possible
places of photon emission are labeled with a  cross.}\label{f2}
\end{figure}
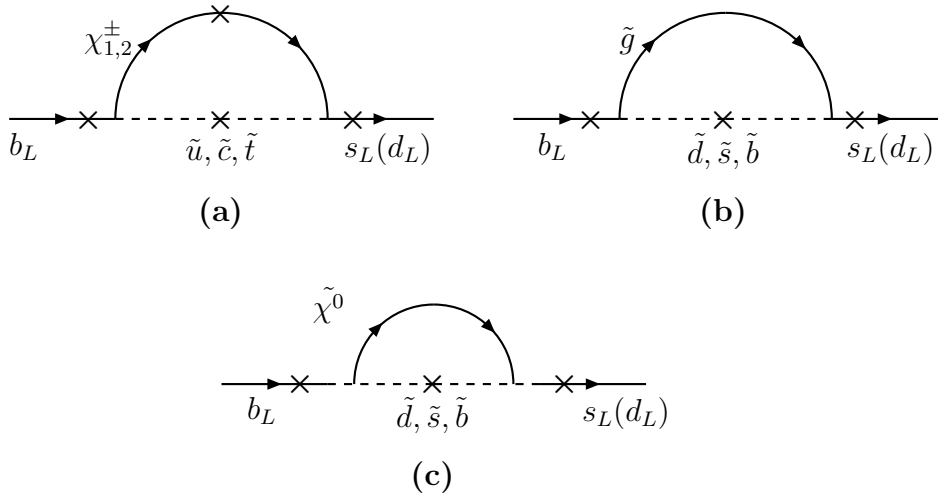

\newpage
\begin{figure}
\begin{center}
\epsfxsize=15.5cm
\epsfysize=12cm
\leavevmode
\mbox{\hskip -2cm}
\epsfbox[20 7 673 377]{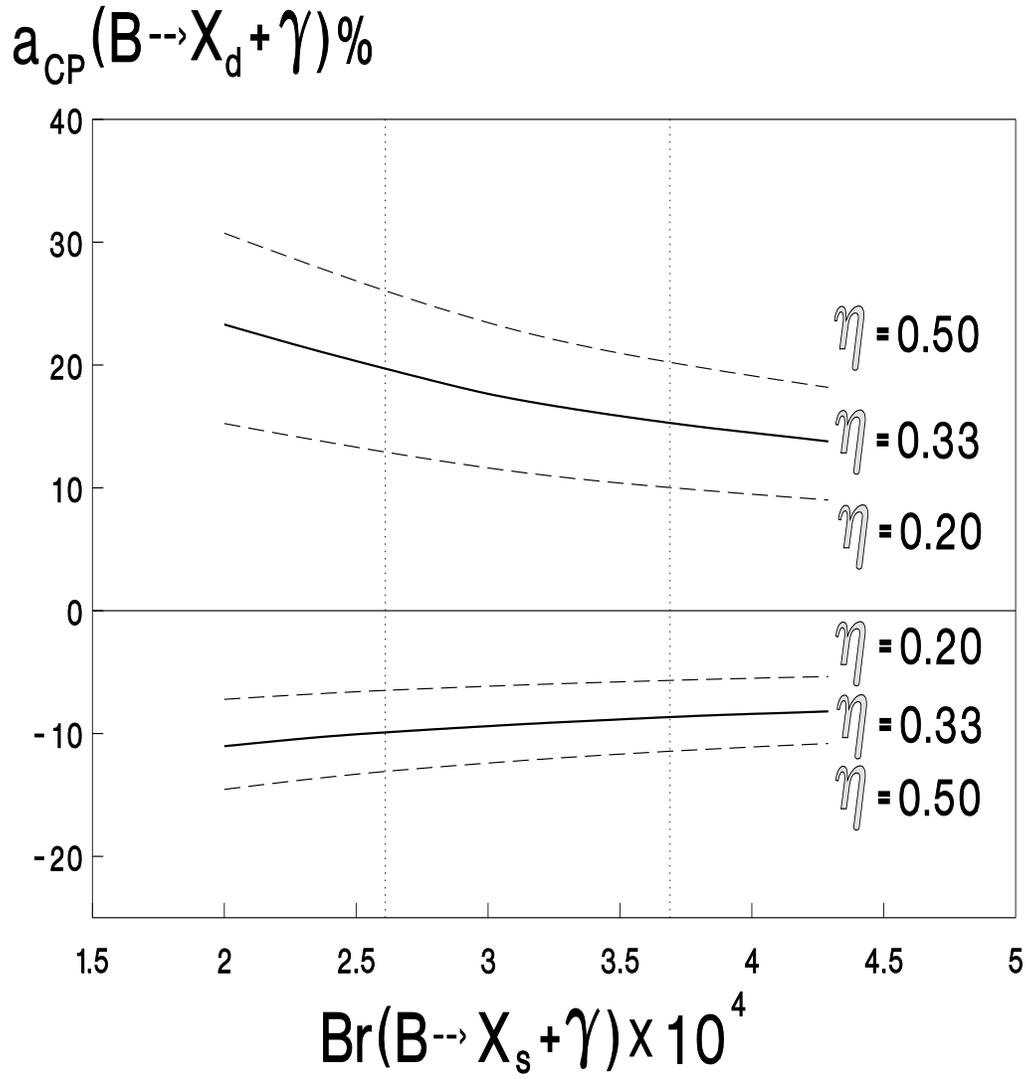}
\end{center}
\caption[]{The dependence of the $a_{CP}(B\to X_d\gamma)$ from 
$BR(B\to X_s\gamma)$ for the values $\eta=0.20,0.33,0.50$ and
$\rho=0.11$, $\mu=2.5GeV$.
The CLEO bounds are shown.}
\end{figure}

\newpage
\begin{figure}
\begin{center}
\epsfxsize=15.5cm
\epsfysize=12cm
\leavevmode
\mbox{\hskip -2cm}
\epsfbox[20 7 673 377]{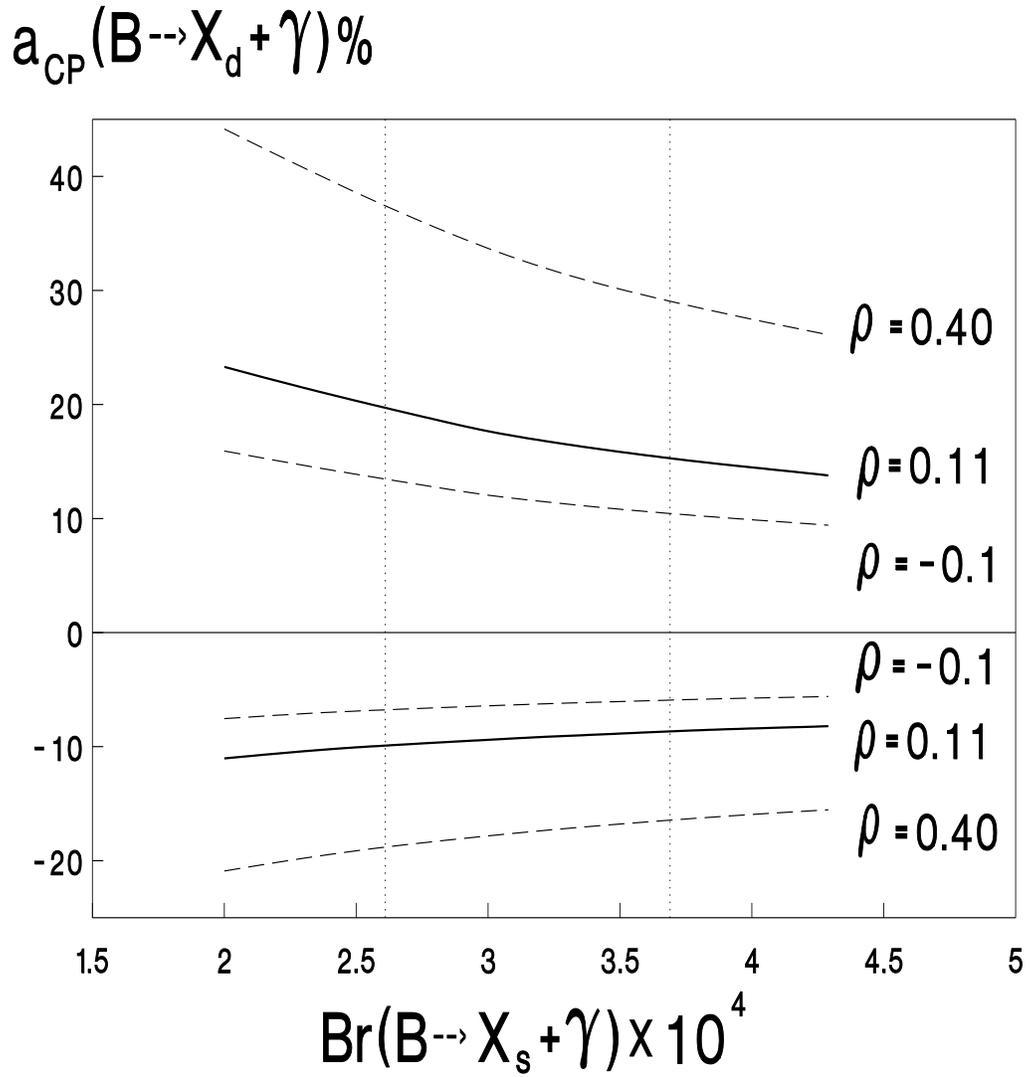}
\end{center}
\caption[]{The dependence of the $a_{CP}(B\to X_d\gamma)$ from 
$BR(B\to X_s\gamma)$ for the $\rho=-0.10,0.11,0.40$ and
$\eta=0.33$, $\mu=2.5GeV$. The CLEO bounds are shown.}
\end{figure}
\end{document}